\documentclass{INTERSPEECH2023}

\interspeechcameraready

\usepackage{amsmath,graphicx}
\usepackage{cite,bm}
\usepackage{amssymb}
\usepackage{amsfonts}
\usepackage{comment}
\usepackage{amsmath,mathtools}
\usepackage{booktabs}
\usepackage[noend]{algpseudocode}
\usepackage{flushend}
\usepackage{siunitx}
\usepackage{xcolor}
\usepackage{arydshln}
\usepackage{url}
\AtBeginDocument{
  \abovedisplayskip     =0.3\abovedisplayskip
  \abovedisplayshortskip=0.3\abovedisplayshortskip
  \belowdisplayskip     =0.3\belowdisplayskip
  \belowdisplayshortskip=0.3\belowdisplayshortskip
}

\def\L{{\mathcal L}}
\def\T{{\mathcal T}}
\def\S{{\mathcal S}}

\def\R{{\mathbb R}}

\title{Remixing-based Unsupervised Source Separation from Scratch}

\name{Kohei Saijo$^1$, Tetsuji Ogawa$^1$}
\address{
  $^1$Department of Communications and Computer Engineering, Waseda University, Tokyo, Japan
}
\email{saijo@pcl.cs.waseda.ac.jp}

\begin{document}

\maketitle

\begin{abstract}

We propose an unsupervised approach for training separation models from scratch using \textit{RemixIT} and \textit{Self-Remixing}, which are recently proposed self-supervised learning methods for refining pre-trained models.
They first separate mixtures with a teacher model and create \textit{pseudo-mixtures} by shuffling and remixing the separated signals.
A student model is then trained to separate the pseudo-mixtures using either the teacher's outputs or the initial mixtures as supervision. 
To refine the teacher's outputs, the teacher's weights are updated with the student's weights.
While these methods originally assumed that the teacher is pre-trained, we show that they are capable of training models from scratch. We also introduce a simple remixing method to stabilize training.
Experimental results demonstrate that the proposed approach outperforms mixture invariant training, which is currently the only available approach for training a monaural separation model from scratch.

\end{abstract}
\section{Introduction}

Remarkable progress has been made in source separation thanks to advances in neural networks~\cite{dc, pit, convtasnet}.
Typically, models are trained with supervised learning, exploiting a large number of the mixture and ground truth pairs.
However, obtaining these pairs in real-world environments is challenging, so they are generally synthesized using simulation toolkits~\cite{pra}.
Unfortunately, the separation performance in real environments is often degraded due to mismatches with real-recorded mixtures, such as differences in reverberation, noise type, or channel mismatch.

To address this issue, unsupervised learning methods directly train models using real-recorded mixtures without ground truths~\cite{drude_unsup, spatial_loss}.
In monaural source separation, mixture invariant training (MixIT)~\cite{mixit} has been highly successful in a variety of separation tasks by training models to separate a mixture of mixtures (MoM), which is created by summing up multiple mixtures~\cite{efficient_mixit, bird_mixit, adapting_mixit}.
However, due to the mismatch between MoMs and the actual mixtures in that MoMs contain more sources, models often suffer from the \textit{over-separation} problem.

To address the over-separation issue, several attempts have been made to refine MixIT pre-trained models in an unsupervised manner.
Teacher-student MixIT~\cite{tsmixit} trains a new student model with fewer output channels, using the MixIT pre-trained model as a teacher.
On the other hand, RemixIT~\cite{remixit, remixit_journal} trains a student model to estimate the teacher's outputs from a \textit{pseudo-mixture} generated by shuffling and remixing the teacher's outputs.
While Teacher-student MixIT uses a static teacher, RemixIT iteratively updates the teacher's weights using the student's weights, leading to improved performance.
Self-Remixing~\cite{selfremixing} has a similar training process to RemixIT but uses initial mixtures as supervision to ensure training stability.

In contrast to the multi-stage training methods that refine MixIT pre-trained models, we propose a novel approach to train models from scratch using RemixIT and Self-Remixing.
While refining pre-trained models has shown success, training from scratch with them has not been explored because of the anticipated high level of distortion in pseudo-mixtures generated by randomly initialized models, as well as the possibility of falling into trivial solutions that do not separate sources.
However, our empirical findings suggest that pseudo-mixtures generated by randomly initialized models can be treated as MoMs, and the training can be done through MixIT-like optimization problems, indicating that RemixIT and Self-Remixing can be effective even when starting from scratch.
Furthermore, we introduce a carefully designed remixing algorithm to overcome the trivial solution issue.
At the start of training, there is a significant mismatch between pseudo-mixtures and actual mixtures since pseudo-mixtures become MoMs.
However, such a mismatch gradually decreases as the teacher model is refined, leading to better performance than using static MoMs.

Our main contributions can be summarized as follows:
\textit{i)} We show how to train models from scratch using RemixIT and Self-Remixing, without the need for pre-training.
Additionally, we propose a remixing method to enhance training stability.
\textit{ii)} We investigate the impact of the remixing algorithm on the separation performance, which has received little attention until now.
Our analysis reveals that the design of the remixing algorithm plays a crucial role in the final separation performance, especially regarding the word error rate.

\section{Methods}
\vspace{-1mm}

\begin{figure*}[t]
\centering
\centerline{\includegraphics[width=0.7\linewidth]{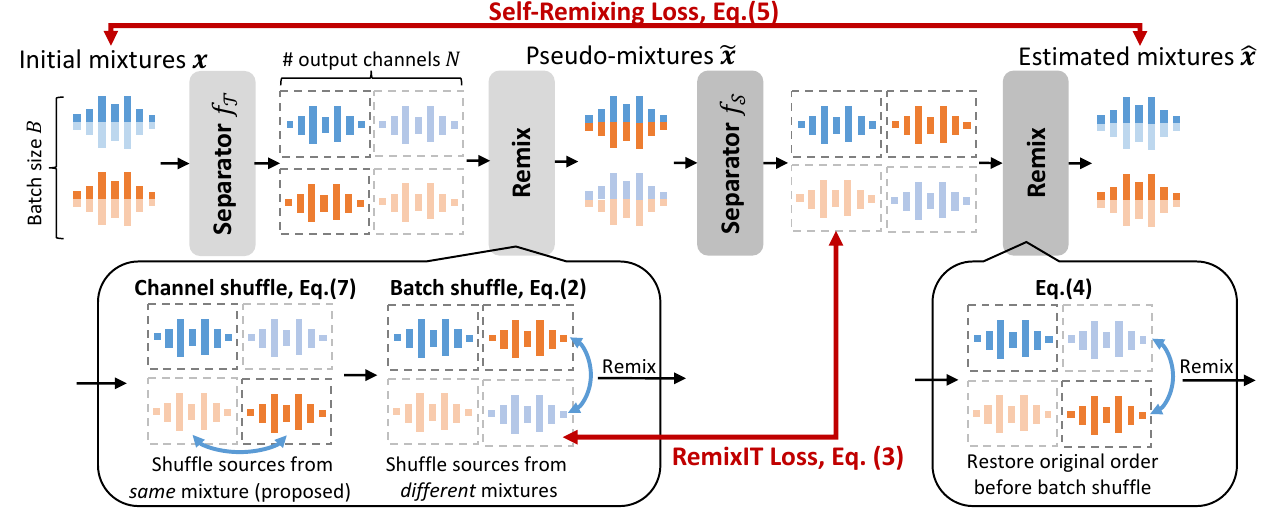}}
\vspace{-2.5mm}
\caption{
    Overview of RemixIT and Self-Remixing.
    We first make pseudo-mixtures by separating mixtures with $f_\T$ and remixing outputs.
    We then train $f_\S$ to separate pseudo-mixtures using outputs of $f_\T$ or initial mixtures as supervision.
    Channel shuffle (bottom left) is introduced to stabilize Self-Remixing.
    Note that they can be applied regardless of $B$ and $N$.
}
\label{fig: overview}
\vspace{-4.5mm}
\end{figure*}

Our proposal is to train separation models from scratch in an unsupervised manner using RemixIT or Self-Remixing.
We start by providing an overview of related methods, including MixIT~\cite{mixit} and mixture permutation invariant training (MixPIT)~\cite{mixpit}.
Then, we delve into the details of RemixIT and Self-Remixing and explain how they work from scratch.
To enhance the training stability, we introduce a simple remixing method.

Let us denote a mini-batch of $B$ mixtures as $\bm{x}\in\mathbb{R}^{B \times T}$, where $T$ denotes the number of samples in the time domain.
Each mixture in the batch, denoted as $x_b$ $(b\!=\!1,\dots,B)$, may contain up to $K$ sources.
To define a separator with $N$ output channels, we use $f_\S$ with its parameters represented by $\bm{\theta}_\S$.

\vspace{-1.5mm}
\subsection{MixIT and MixPIT}
\vspace{-1mm}

To generate a mixture of mixtures (MoM) $\bar{x}$, MixIT takes a mini-batch $\bm{x}$ and adds sets $B'$ mixtures together, denoted as $x_1, \ldots, x_{B'}$.
The resulting $\bar{x}$ is then fed into $f_\S$ to produce separated signals $\hat{\bm{s}} \in\R^{N \times T}$ (${N} \geq B'K$).
The MixIT loss is then computed between these separated signals and the individual mixtures, as described in~\cite{mixit}:
\begin{align}
  \label{eqn:mixit_loss}
    \L_{\rm{MixIT}} = \min_{\bm{A}} \sum\nolimits_{b'=1}^{B'} {\L(x_{b'}, [\bm{A}\hat{\bm{s}}]_{b'})},
\end{align}
where $\L$ is a loss function and $\bm{A}\in\mathbb{B}^{{B'}\times{N}}$ assigns each $\hat{s}_{n}$ to the original mixtures.
$B'$ is typically set to two.
While MixIT has made unsupervised monaural source separation feasible, the models often encounter an over-separation issue due to the fact that MoMs contain more sources than individual mixtures.

MixPIT is a recent development that aims to address the over-separation problem.
MixPIT creates MoMs by summing the same number of mixtures as the number of sources in individual mixtures ($K$) and trains the models to estimate the individual mixtures directly (i.e., $B'=N=K$).
Despite underperforming MixIT, as reported in~\cite{mixpit}, the success of MixPIT implies that RemixIT and Self-Remixing can also be used effectively from scratch (as detailed in Section \ref{ssec: proof}).

\vspace{-1.5mm}
\subsection{RemixIT and Self-Remixing}
\label{ssec:remixit_and_selfremixing}
\vspace{-1mm}

An overview of RemixIT and Self-Remixing is provided in Fig.~\ref{fig: overview}.
RemixIT involves a teacher model, denoted as $f_\T$, that generates pseudo-mixtures from observed mixtures, and a student model, denoted as $f_\S$, that is trained to separate these pseudo-mixtures.
The student model is trained via gradient descent using the teacher's outputs as supervision, while the teacher's parameters $\bm{\theta}_\T$ are updated using the student's parameters $\bm{\theta}_\S$.
Self-Remixing follows a similar training procedure as RemixIT but uses the initial mixtures as supervision.

The teacher $f_\T$ first separates the mixtures $\bm{x}$ into sources: $\tilde{\bm{s}}=f_\T(\bm{x};\bm{\theta}_\T)\in\mathbb{R}^{B \times N \times T}$.
To ensure that the sources add up to the initial mixtures, $\sum\nolimits_{n=1}^{N}{\tilde{s}_{b,n}} = x_b$, we enforce mixture consistency (MC)~\cite{mixconsis}.
Next, we shuffle the sources within a batch in each output channel $n$ with an $B \times B$ permutation matrix $\bm{\Pi}_n$, and make pseudo-mixtures $\tilde{\bm{x}}$ by adding them together:
\begin{align}
  \label{eqn:remix}
    \tilde{x}_{b} = \sum\nolimits_{n=1}^{N} {\tilde{s}_{b,n}}^{(\bm{\Pi})}, ~~~~~~{\tilde{s}_{b,n}}^{(\bm{\Pi})} = [{{\bm{\Pi}}_n}{{\tilde{\bm{s}}}^{\top}_n}]_{b},
\end{align}
where we define $^{\top}$ as the transpose of the first and the second dimension ($\R^{B \times N \times T} \rightarrow \R^{N \times B \times T}$) and ${\tilde{\bm{s}}}^{\top}_n \triangleq [\tilde{s}_{1,n}, \ldots, \tilde{s}_{B,n}]$.
The process described in Eq.~(\ref{eqn:remix}) is referred to as \textit{batch shuffle}.
After remixing, $f_\S$ separates the pseudo-mixtures: $\hat{\bm{s}}=f_\S(\tilde{\bm{x}};\bm{\theta}_{\S})$.
The loss for RemixIT is computed in a permutation-invariant way~\cite{pit} between the separated signals of $f_\T$ and $f_\S$:
\begin{align}
  \label{eqn:remixit_loss}
    \L_{\rm{RemixIT}}^{(b)} = \min_{\bm{P}} \frac{1}{N} \sum\nolimits_{n=1}^{N} {\L({\tilde{s}}^{(\bm{\Pi})}_{b,n}, [\bm{P}_{b}\hat{\bm{s}}_{b}]_{n})},
\end{align}
where ${\bm{P}_{b}}$ is an $N \times N$ permutation matrix.
We then align $\hat{\bm{s}}_{b}$ in the same order as $\tilde{\bm{s}}^{(\bm{\Pi})}_{b}$ using the optimal permutation matrix $\bar{\bm{P}}_{b}$: ${\hat{\bm{s}}^{(\bar{\bm{P}})}}_{b} = \bar{\bm{P}}_{b}\hat{\bm{s}}_{b}$.
Next, we restore the order of sources to their original sequence before the batch shuffle in Eq.~(\ref{eqn:remix}):
\begin{align}
  \label{eqn:shuffle_back}
    {{\hat{s}_{b,n}}^{(\bm{\Pi}^{-1})}} = [{{\bm{\Pi}}^{-1}_n}{{\hat{\bm{s}}}^{(\bar{\bm{P}})\top}_n}]_{b},
\end{align}
where ${{\hat{\bm{s}}}^{(\bar{\bm{P}})\top}_n} \triangleq [\hat{s}^{(\bar{\bm{P}})}_{1,n}, \ldots, \hat{s}^{(\bar{\bm{P}})}_{B,n}]$.
The Self-Remixing loss is computed using the initial mixtures as supervision:
\begin{align}
  \label{eqn:selfremixing_loss}
  \L_{\rm{Self-Remixing}}^{(b)} = \L\left(x_b, \sum\nolimits_{n=1}^{N} {{\hat{s}_{b,n}}^{(\bm{\Pi}^{-1})}}\right).
\end{align}

While $f_\S$ is trained via gradient descent using Eqs.(\ref{eqn:remixit_loss}) or (\ref{eqn:selfremixing_loss}), the teacher's parmeters $\bm{\theta}_\T$ are updated using the student's parameters $\bm{\theta}_\S$.
In this work, we update $\bm{\theta}_\T$ in the exponential moving average (EMA) fashion at every epoch end:
\begin{align}
  \label{eqn:teacher_update}
    \bm{\theta}_{\T}^{(j+1)} = \alpha\bm{\theta}_{\T}^{(j)} + (1-\alpha)\bm{\theta}_{\S}^{(j)},
\end{align}
where $\alpha\in[0,1]$ and $j$ is the epoch index.
We set $\alpha=0.8$.

\vspace{-1.5mm}
\subsection{Why do RemixIT/Self-Remixing work from scratch?}
\label{ssec: proof}
\vspace{-1mm}

In this work, we demonstrate that RemixIT and Self-Remixing can work even when $f_\T$ is randomly initialized, contrary to the original assumption that $f_\T$ is pre-trained.

We analyzed the output of a model initialized randomly\footnote{We used the default initialization strategy of Pytorch 1.12.1~\cite{pytorch}.} using the Conformer architecture described in Section \ref{ssec: separation_model}, and noisy two-speaker mixtures described in Section \ref{ssec: datasets}.
Our empirical results indicated that the outputs after enforcing MC tended to be similar to the input mixture, albeit with different scales.
Specifically, the average scale-invariant signal-to-distortion ratio (SISDR)~\cite{sisdr} of the outputs compared to the input mixture was 18.4 dB, indicating that they were somewhat distorted, but the deviation from the initial mixtures was small.

Based on the assumption that the outputs of $f_\T$ can be considered as mixtures, the pseudo-mixtures created by remixing them can be treated as MoMs.
Given this assumption, the loss function used in RemixIT is the same as MixPIT's because RemixIT trains the models to separate MoMs into each mixture.
In addition, Self-Remixing also promotes the separation of MoMs to reconstruct the initial mixtures.
Therefore, the objectives of RemixIT and Self-Remixing are similar to MixPIT's.
These analyses suggest that RemixIT and Self-Remixing can be effective even when starting from random initialization.

Note that while our analysis was conducted using the Conformer architecture, depending on the architecture, there may be cases where the outputs of $f_\T$ deviate from the input mixtures and the above assumption does not hold.
However, it is possible to ensure that the assumption holds, e.g., by initializing the output layer of $f_\T$ with zeros.
This results in all outputs having the same values (e.g., simply zeros or 0.5 with the sigmoid activation) and the MC layer scales the sources to be $\frac{1}{N}$ mixtures, so pseudo-mixtures become MoMs.
While models were randomly initialized in this work, we confirmed that RemixIT and Self-Remixing still worked when we initialized the final linear layer of the Conformer with zeros.

\vspace{-1.5mm}
\subsection{Design of Remixing}
\label{ssec: remixing_algorithm}
\vspace{-1mm}

We introduce our proposed remixing algorithm which has been specifically designed to facilitate training from scratch.

\noindent\textbf{Channel shuffle (CS)}:
In the previous subsection, we discussed how RemixIT and Self-Remixing can operate without pre-training.
However, we discovered that Self-Remixing can encounter a trivial solution in that sources are not separated at all because the Self-Remixing loss can be minimized even when the model does not separate sources.
Specifically, the model outputs the input mixture from one output channel and zeros from the others, e.g., $\tilde{s}_{b,1}=x_b$ and $\tilde{s}_{b,n \neq 1}=0$.
Since the batch shuffle (as seen in Eq.~(\ref{eqn:remix})) is conducted in each output channel, the model may easily fall into such a trivial solution.
To address this problem, we propose a solution called \textit{channel shuffle} (CS), which shuffles sources that are separated from the same mixture before the batch shuffle:
\begin{align}
  \label{eqn:shuffle_source_dim}
    {\tilde{\bm{s}}}_{b} \leftarrow \bm{\Lambda}_{b}{\tilde{\bm{s}}_{b}} \in\mathbb{R}^{N \times T},
\end{align}
where $\bm{\Lambda}_{b}$ is an ${N \times N}$ permutation matrix.
We show this simple method greatly improves the stability of Self-Remixing.

\noindent\textbf{Avoiding remixing of sources from the same mixture}:
When the permutation matrix $\bm{\Pi}_{n}$ in Eq.(\ref{eqn:remix}) is randomly set, pseudo-mixtures sometimes contain the sources separated from the same mixture.
However, such pseudo-mixtures interfere with training when applying RemixIT from scratch.
Let us consider the case where $f_\T$ with $N=3$ is randomly initialized and the pseudo-mixture is composed of the following three sources $\tilde{s}_1=\frac{1}{3}x_1, \tilde{s}_2=\frac{1}{3}x_2, \tilde{s}_3=\frac{1}{3}x_2$ (i.e., $\tilde{x} = \frac{1}{3}x_1 + \frac{2}{3}x_2$).
In this case, RemixIT loss encourages $f_\S$ to separate $\frac{2}{3}x_2$ into two $\frac{1}{3}x_2$; however, this objective does not necessarily improve separation performance.
As a result, when applying RemixIT, we opt to select the permutation matrix $\bm{\Pi}_{n}$ under the condition that sources from the same mixture are not remixed.
Note that Self-Remixing is still effective even when the sources from the same mixture are remixed because it encourages $f_\S$ to separate $\tilde{x}$ into $\frac{1}{3}x_1$ and $\frac{2}{3}x_2$ to reconstruct initial mixtures, $x_1$ and $x_2$.

\vspace{-1.5mm}
\section{Experiments}

\vspace{-1.5mm}
\subsection{Datasets}
\label{ssec: datasets}
\vspace{-1mm}
\noindent\textbf{WSJ-mix}:
We synthesized two-speaker mixtures with reverberation and noise using speeches from WSJ0~\cite{wsj0} and WSJ1~\cite{wsj1} and noises from CHiME3~\cite{chime3} at a sampling rate of 8\si{\kilo\hertz}.
The configuration was similar to that of SMS-WSJ~\cite{smswsj} but with changes in reverberation times, and noise type and level.
Pyroomacoustics~\cite{pra} was used for simulation, and the reverberation times were chosen from \SIrange{0.2}{1.0}{\second}.
Additionally, all source image locations were jittered by up to 8cm to avoid the sweeping echo problem~\cite{sweeping_echo}.
Note that the condition is different from our previous paper~\cite{selfremixing}.
The SNR of the noise ranged between \SIrange{10}{20}{\decibel}.
The dataset consisted of 33561 ($\sim$87.4h), 982 ($\sim$2.5h), and 1332 ($\sim$3.6h) mixtures for the training, validation, and test sets, respectively.
To evaluate word error rates (WERs), we used the ASR backend provided in~\cite{smswsj}.

\noindent\textbf{FUSS}:
We utilized the anechoic version of the free universal sound separation (FUSS) dataset~\cite{fuss} to evaluate the performance in USS tasks~\cite{universal_sound_separation}.
The dataset comprised mixtures with one to four sources, drawn from 357 classes of audio sources.
All mixtures were ten seconds long and sampled at 16\si{\kilo\hertz}.
The training, validation, and test sets contained 20000, 1000, and 1000 mixtures, respectively.

\vspace{-1.5mm}
\subsection{Separation model}
\label{ssec: separation_model}
\vspace{-1mm}

As the separation model, we utilized Conformer~\cite{conformer} (implemented based on \cite{libricss_conformer}).
The model contained about 21.6M parameters and was composed of 16 Conformer encoder layers with four attention heads, 256 attention dimensions, and 1024 feed-forward network dimensions.
We replaced the batch normalization~\cite{batchnorm} with the group normalization~\cite{groupnorm} with eight groups.
It took log-magnitude (or magnitude) spectrograms in the STFT domain as inputs and produced real-valued TF masks in WSJ-mix (FUSS).
The FFT size was 512 and the window size and the hop length were 400 and 160, respectively.

\vspace{-1.5mm}
\subsection{Compared methods}
\vspace{-1mm}

We compared the performance of several methods, including \textbf{MixIT}, \textbf{MixIT+Sparsity}, \textbf{RemixIT}, \textbf{Self-Remixing}, and their supervised counterparts, \textbf{Sup.~RemixIT} and \textbf{Sup.~Self-Remixing}.
MixIT is implemented based on~\cite{asteroid}.
In MixIT and MixIT+Sparsity, $N$ was set to six in WSJ-mix and eight in FUSS, and we ensured MC.
In the other methods, $N$ was set to three in WSJ-mix and four in FUSS.
MixIT+Sparsity uses the sparsity loss (Eq.~(6) in~\cite{efficient_mixit}) that promotes sparsity of the outputs and prevents over-separation.
Note that the sparsity loss does not necessarily improve separation performance, as demonstrated in~\cite{efficient_mixit}.
We trained RemixIT and Self-Remixing from scratch using Eqs.~(\ref{eqn:remixit_loss}) and (\ref{eqn:selfremixing_loss}), respectively. For the supervised versions, we used ground-truth signals instead of the outputs of $f_\T$ to create pseudo-mixtures.

We excluded MixPIT from the baseline because prior studies have demonstrated that MixIT outperforms MixPIT~\cite{mixpit}.
Additionally, we did not consider remix-cycle-consistent learning~\cite{saijo_rccl, saijo_rccl2}, which is similar to Self-Remixing but shares parameters between $\bm{\theta}_\T$ and $\bm{\theta}_\S$ and computes the gradient through two separation and remixing processes because we did not achieve good results when training it from scratch.

\vspace{-1.5mm}
\subsection{Training details}
\vspace{-1mm}

As the signal-level loss function $\mathcal{L}$, we used the negative thresholded SNR between the reference $y$ and the estimate $\hat{y}$:
\begin{align}
  \label{eqn:snr_loss}
    \L(y, \hat{y}) = -10\log_{10}{\frac{||y||^2}{||y-\hat{y}||^2 + \tau||y||^2}},
\end{align}
where $\tau = 10^{-3}$ is a threshold that clamps the SNR at 30~\si{\decibel}.
In supervised methods, we used the SNR loss that can handle zero-references (Eq.~(2) of~\cite{fuss}), instead of Eq.~(\ref{eqn:snr_loss}).

When training with WSJ-mix, the batch size was 32 and the input was 7 seconds long.
Models were trained for 600 epochs.
When training with FUSS, the batch size was 16 and the input was 10 seconds long.
Models were trained for 400 epochs.
In both experiments, we used the AdamW optimizer~\cite{adamw} with the weight decay of 1e-2.
The learning rate was increased linearly from 0 to 2e-4 over the first 5000 training steps, kept constant at 2e-4 for 100 epochs, and then decayed by 0.98 for every three epochs until it reached 2e-5.
The gradients were clipped with a maximum norm of 5.
In MixIT+Sparsity, we initially trained the model for 450 (300) epochs using only the MixIT loss.
We then fine-tuned the model with the sparsity loss for 150 (100) epochs in WSJ-mix (FUSS), with the weight of sparsity loss set to 4 (23).
In supervised methods, all-zero pseudo-mixtures were occasionally generated, but we did not use them for training.

To normalize each mixture, we subtracted its mean and divided it by its standard deviation.
For MoM $\bar{x}$ in MixIT, we followed the same process.
Although \cite{selfremixing} reported the instability of RemixIT in source separation tasks, we found that normalization improves the stability of the model. As a result, we successfully trained models on reverberant noisy two-speaker mixtures similar to the dataset used in \cite{selfremixing}.
It is worth noting that we only normalized the initial mixtures $x$ and did not normalize the pseudo-mixtures $\tilde{x}$ because normalizing $\tilde{x}$ resulted in lower performance in our experiments.

\vspace{-1.5mm}
\subsection{Results on WSJ-mix}
\vspace{-1mm}
\begin{table}[t]
\begin{center}
\caption{
    Evaluation results on WSJ-mix test set.
}
\vspace{-3.5mm}
\label{table:results_wsj}
\resizebox{0.85\linewidth}{!}{
\begin{tabular}{lrrcc}
\toprule
{Method} & {SISDR} & {STOI} & {PESQ} & {WER}  \\

\midrule
    Unprocessed &-0.4  &0.684 &1.82 &82.9\% \\ 

\midrule
    \texttt{A1} MixIT          &8.8 &0.847 &2.54 &42.3\% \\
    \texttt{A2} ~~~+ Sparsity  &8.6 &0.843 &2.51 &44.8\% \\
    \texttt{A3} RemixIT w/o CS &\bf{10.8} &\bf{0.890} &\bf{2.84} &47.4\% \\
    \texttt{A4} RemixIT        &10.3 &0.878 &2.75 &43.3\% \\
    \texttt{A5} Self-Remixing  &10.3 &0.877 &2.69 &50.1\% \\
    \texttt{A6} Self-Remixing$^\dagger$ &10.3 &0.878 &2.74 &\bf{39.7}\% \\
    
\midrule
    Sup. RemixIT w/o CS &10.9  &0.896 &3.01 &30.9\% \\ 
    Sup. RemixIT        &10.6  &0.889 &2.93 &33.4\% \\ 
    Sup. Self-Remixing  &10.6  &0.889 &2.93 &33.7\% \\

\bottomrule

\end{tabular}}
\end{center}
\vspace{-7mm}
\end{table}

\begin{figure}[t]
\centering
\centerline{\includegraphics[width=0.79\linewidth]{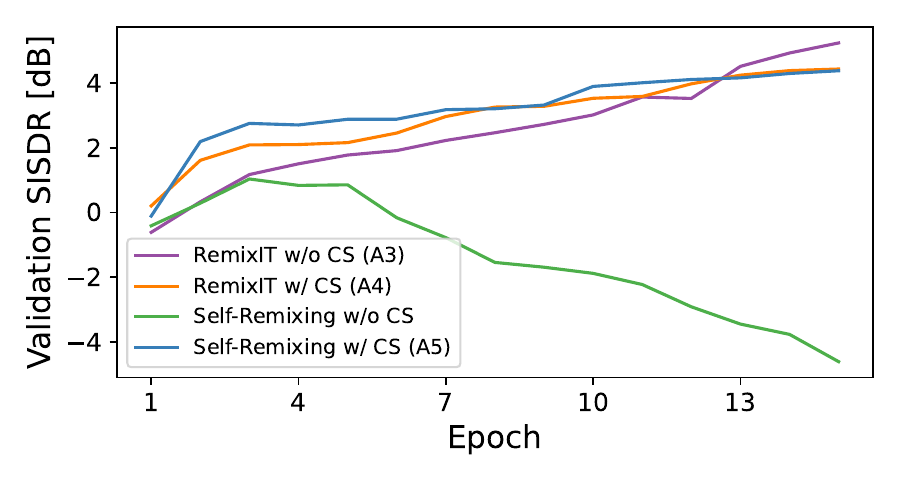}}
\vspace{-3.75mm}
\caption{
    Validation SISDR of $f_\S$ of RemixIT and Self-Remixing in early stage of training.
}
\label{fig: training_curve}
\vspace{-6mm}
\end{figure}

The results for SISDR~\cite{fast_bss_eval}, short-time objective intelligibility (STOI)\cite{stoi}, perceptual evaluation of speech quality (PESQ)~\cite{pesq}, and WER on WSJ-mix are shown in Table~\ref{table:results_wsj}.
To evaluate the model, we used the averaged parameters from five checkpoints that yielded the highest SISDR on the validation set.
Note that $\dagger$ denotes instances where we allowed remixing of sources from the same mixture (e.g., Self-Remixing$^\dagger$ in \texttt{A6}).

The results show that both RemixIT and Self-Remixing performed well from scratch and outperformed MixIT in terms of speech metrics.
However, the WERs of \texttt{A3-A5} were worse than that of MixIT.
As discussed in~\cite{selfremixing}, remixing-based methods are trained with distorted pseudo-mixtures, which can result in models outputting distorted signals.
This implies that the distortion caused by using pseudo-mixtures is more harmful to speech recognition than that caused by using MoMs for training.
Despite this, the substantial improvement in the WER from \texttt{A5} to \texttt{A6} suggests that allowing remixing of sources from the same mixture can alleviate distortion in separated signals.

Fig.~\ref{fig: training_curve} displays the validation SISDR in the early stage of training.
It has been confirmed that CS stabilizes Self-Remixing.
This outcome, along with the observed WER improvement in \texttt{A6}, highlights the significance of the remixing algorithm.
It impacts both the final separation performance and training stability.
In future work, we plan to explore a more thoughtfully designed remixing approach to enhance the performance of both Self-Remixing and RemixIT.

\vspace{-1.5mm}
\subsection{Results on FUSS}
\vspace{-1mm}

\begin{table}[t]
\begin{center}
\caption{
    Evaluation results on FUSS test set. 
    Checkpoints that gave best MSi (\texttt{B*}) and TRF (\texttt{C*}) were evaluated. 
}
\vspace{-3.75mm}
\label{table:results_fuss}
\resizebox{0.95\linewidth}{!}{
\begin{tabular}{lrrrrrr}
\toprule
{Method} & {1S} & {2Si} & {3Si} & {4Si} & {MSi} & {TRF}  \\

\midrule
    \texttt{B1} MixIT          &9.5 &9.8 &14.4 &15.9 &13.2 &12.2 \\
    \texttt{B2} ~~~+ Sparsity  &15.2 &12.3 &15.5 &\bf{16.6} &14.7 &14.8 \\
    \texttt{B3} RemixIT        &42.1 &12.3 &11.7 &8.2 &10.8 &19.1 \\
    \texttt{B4} Self-Remixing  &29.4 &14.7 &\bf{16.2} &13.6 &\bf{14.8} &18.7 \\
    \texttt{B5} Self-Remixing$^\dagger$ &31.9 &14.9 &15.9 &13.7 &\bf{14.8} &19.4 \\
    Sup. Self-Remixing  &32.9 &14.1 &15.8 &13.5 &14.5 &19.4 \\ 
    \midrule
    
    \texttt{C1} MixIT          &9.5 &9.9 &14.3 &15.8 &13.2 &12.2 \\
    \texttt{C2} ~~~+ Sparsity  &18.0 &12.3 &15.5 &16.5 &14.7 &15.6 \\
    \texttt{C3} RemixIT        &\bf{63.6} &4.6  &2.1  &0.1  &2.4 &18.6 \\
    \texttt{C4} Self-Remixing  &34.4 &14.8 &15.9 &13.1 &14.6 &19.8 \\
    \texttt{C5} Self-Remixing$^\dagger$ &34.9 &\bf{15.0} &15.6 &13.4 &14.7 &\bf{20.0} \\
    Sup. Self-Remixing  &49.2 &13.7 &14.7 &12.0 &13.5 &23.0 \\ 
      
\bottomrule

\end{tabular}}
\end{center}
\vspace{-8mm}
\end{table}

The evaluation results on the FUSS test set are listed in Table~\ref{table:results_fuss}, which includes the following evaluation metrics~\cite{efficient_mixit}: \textbf{1S}, representing the SISDR for single-source mixtures; \textbf{$\bm{k}$Si}, representing the SISDR improvement (SISDRi) for $k$-source mixtures ($k=2,3,4$); \textbf{MSi}, representing the average SISDRi for multi-source mixtures; and \textbf{TRF}, representing the total reconstruction fidelity with average performance on all test data.
For evaluation, we utilized the averaged model parameters from the five checkpoints that resulted in the highest MSi (\texttt{B*}) and TRF (\texttt{C*}) on the validation set.
It is worth noting that, unlike~\cite{fuss}, we did not discard estimate-reference pairs with low-energy estimated sources following~\cite{sudormrf}, as we believe that errors in low-energy estimates should be taken into account.

The results indicate that Self-Remixing achieved higher TRF than MixIT+Sparsity while providing almost the same MSi.
Unlike MixIT, which suffers from low 1S due to the inclusion of at least two sources in MoMs, Self-Remixing attains high performance in both 1S and MSi by utilizing both single-source and multi-source pseudo-mixtures.

In RemixIT, the separation performance initially improved, but then it deteriorated.
This can be attributed to the fact that many mixtures have fewer sources than the number of output channels, and as a result, $f_\T$ often outputs sources that are close to zero.
In such a case, the RemixIT loss encourages $f_\S$ to output zeros, leading to \textit{under-separation}.
As the teacher's weights are updated with the student's weights, under-separation occurs more frequently in $f_\T$, resulting in high 1S but low MSi.

Interestingly, the unsupervised Self-Remixing resulted in slightly higher MSi than the supervised one.
We attribute this to the over-separation of $f_\T$.
Specifically, when $f_\T$ separates three-source mixtures, for instance, over-separation can lead to the output of four sources instead of the expected three sources and one zero signal.
As a result, the average number of sources contained in pseudo-mixtures becomes higher than that of supervised learning, leading to higher MSi and lower 1S.
\vspace{-1.5mm}
\section{Conclusion}
\vspace{-1.5mm}

We proposed to train separators from scratch with RemixIT and Self-Remixing.
We showed that their training objectives are similar to MixPIT's at the start of the training and thus they are capable of training without any pre-training.
The experiments demonstrated the effectiveness of RemixIT and Self-Remixing even when training from scratch, as well as the importance of remixing method.
In the future, we will explore the applicability of the proposed method to other separation models~\cite{convtasnet, tfgridnet}.

\vspace{-2mm}
\section{Acknowledgments}
\vspace{-1.5mm}
The research was supported by NII CRIS collaborative research program operated by NII CRIS and LINE Corporation.

\bibliographystyle{IEEEtran}
\bibliography{main}

\begin{thebibliography}{10}
\providecommand{\url}[1]{#1}
\csname url@samestyle\endcsname
\providecommand{\newblock}{\relax}
\providecommand{\bibinfo}[2]{#2}
\providecommand{\BIBentrySTDinterwordspacing}{\spaceskip=0pt\relax}
\providecommand{\BIBentryALTinterwordstretchfactor}{4}
\providecommand{\BIBentryALTinterwordspacing}{\spaceskip=\fontdimen2\font plus
\BIBentryALTinterwordstretchfactor\fontdimen3\font minus
  \fontdimen4\font\relax}
\providecommand{\BIBforeignlanguage}[2]{{%
\expandafter\ifx\csname l@#1\endcsname\relax
\typeout{** WARNING: IEEEtran.bst: No hyphenation pattern has been}%
\typeout{** loaded for the language `#1'. Using the pattern for}%
\typeout{** the default language instead.}%
\else
\language=\csname l@#1\endcsname
\fi
#2}}
\providecommand{\BIBdecl}{\relax}
\BIBdecl

\bibitem{dc}
J.~R. Hershey, Z.~Chen, J.~Le~Roux, and S.~Watanabe, ``Deep clustering:
  Discriminative embeddings for segmentation and separation,'' in \emph{Proc.
  ICASSP}, 2016, pp. 31--35.

\bibitem{pit}
D.~Yu, M.~Kolbæk, Z.~H. Tan, and J.~Jensen, ``Permutation invariant training
  of deep models for speaker-independent multi-talker speech separation,'' in
  \emph{Proc. ICASSP}, 2017, pp. 241--245.

\bibitem{convtasnet}
Y.~Luo and N.~Mesgarani, ``Conv-tasnet: Surpassing ideal time--frequency
  magnitude masking for speech separation,'' \emph{IEEE/ACM transactions on
  audio, speech, and language processing}, vol.~27, no.~8, pp. 1256--1266,
  2019.

\bibitem{pra}
R.~Scheibler, E.~Bezzam, and I.~Dokmanić, ``Pyroomacoustics: A python package
  for audio room simulation and array processing algorithms,'' in \emph{Proc.
  ICASSP}, 2018, pp. 351--355.

\bibitem{drude_unsup}
L.~Drude, D.~Hasenklever, and R.~Haeb-Umbach, ``Unsupervised training of a deep
  clustering model for multichannel blind source separation,'' in \emph{Proc.
  ICASSP}, 2019, pp. 695--699.

\bibitem{spatial_loss}
K.~Saijo and R.~Scheibler, ``Spatial loss for unsupervised multi-channel source
  separation,'' in \emph{Proc. {Interspeech}}, 2022, pp. 241--245.

\bibitem{mixit}
S.~Wisdom, E.~Tzinis, H.~Erdogan, R.~Weiss, K.~Wilson, and J.~Hershey,
  ``Unsupervised sound separation using mixture invariant training,''
  \emph{Advances in Neural Information Processing Systems}, vol.~33, pp.
  3846--3857, 2020.

\bibitem{efficient_mixit}
S.~Wisdom, A.~Jansen, R.~J. Weiss, H.~Erdogan, and J.~R. Hershey, ``Sparse,
  efficient, and semantic mixture invariant training: Taming in-the-wild
  unsupervised sound separation,'' in \emph{Proc. WASPAA}, 2021, pp. 51--55.

\bibitem{bird_mixit}
T.~Denton, S.~Wisdom, and J.~R. Hershey, ``Improving bird classification with
  unsupervised sound separation,'' in \emph{Proc. ICASSP}, 2022, pp. 636--640.

\bibitem{adapting_mixit}
A.~Sivaraman, S.~Wisdom, H.~Erdogan, and J.~R. Hershey, ``Adapting speech
  separation to real-world meetings using mixture invariant training,'' in
  \emph{Proc. ICASSP}.\hskip 1em plus 0.5em minus 0.4em\relax IEEE, 2022, pp.
  686--690.

\bibitem{tsmixit}
J.~Zhang, C.~Zorilă, R.~Doddipatla, and J.~Barker, ``{Teacher-Student MixIT
  for Unsupervised and Semi-Supervised Speech Separation},'' in \emph{Proc.
  {Interspeech}}, 2021, pp. 3495--3499.

\bibitem{remixit}
E.~Tzinis, Y.~Adi, V.~K. Ithapu, B.~Xu, and A.~Kumar, ``Continual self-training
  with bootstrapped remixing for speech enhancement,'' in \emph{Proc.
  ICASSP}.\hskip 1em plus 0.5em minus 0.4em\relax IEEE, 2022, pp. 6947--6951.

\bibitem{remixit_journal}
E.~Tzinis, Y.~Adi, V.~K. Ithapu, B.~Xu, P.~Smaragdis, and A.~Kumar, ``Remixit:
  Continual self-training of speech enhancement models via bootstrapped
  remixing,'' \emph{IEEE Journal of Selected Topics in Signal Processing},
  2022.

\bibitem{selfremixing}
K.~Saijo and T.~Ogawa, ``Self-remixing: Unsupervised speech separation via
  separation and remixing,'' \emph{arXiv preprint arXiv:2211.10194}, 2022.

\bibitem{mixpit}
E.~Karamatl{\i} and S.~K{\i}rb{\i}z, ``Mixcycle: Unsupervised speech separation
  via cyclic mixture permutation invariant training,'' \emph{IEEE Signal
  Processing Letters}, 2022.

\bibitem{mixconsis}
S.~Wisdom, J.~R. Hershey, K.~Wilson, J.~Thorpe, M.~Chinen, B.~Patton, and R.~A.
  Saurous, ``Differentiable consistency constraints for improved deep speech
  enhancement,'' in \emph{Proc. ICASSP}.\hskip 1em plus 0.5em minus 0.4em\relax
  IEEE, 2019, pp. 900--904.

\bibitem{pytorch}
A.~Paszke \emph{et~al.}, ``Pytorch: An imperative style, high-performance deep
  learning library,'' \emph{Advances in neural information processing systems},
  vol.~32, 2019.

\bibitem{sisdr}
J.~Le~Roux, S.~Wisdom, H.~Erdogan, and J.~R. Hershey, ``Sdr--half-baked or well
  done?'' in \emph{Proc. ICASSP}, 2019, pp. 626--630.

\bibitem{wsj0}
J.~S. Garofolo \emph{et~al.}, \emph{{CSR-I} ({WSJ0}) Complete {LDC93S6A}},
  Linguistic Data Consortium, Philadelphia, 1993, web Download.

\bibitem{wsj1}
{Linguistic Data Consortium, and NIST Multimodal Information Group},
  \emph{{CSR-II} ({WSJ1}) Complete {LDC94S13A}}, Linguistic Data Consortium,
  Philadelphia, 1994, web Download.

\bibitem{chime3}
J.~Barker, R.~Marxer, E.~Vincent, and S.~Watanabe, ``The third `{CHiME}' speech
  separation and recognition challenge: Dataset, task and baselines,'' in
  \emph{Proc. ASRU}, 2015, pp. 504--511.

\bibitem{smswsj}
L.~Drude, J.~Heitkaemper, C.~Boeddeker, and R.~Haeb-Umbach, ``Sms-wsj:
  Database, performance measures, and baseline recipe for multi-channel source
  separation and recognition,'' in \emph{arXiv preprint arXiv:1910.13934},
  2019.

\bibitem{sweeping_echo}
E.~De~Sena, N.~Antonello, M.~Moonen, and T.~van Waterschoot, ``On the modeling
  of rectangular geometries in room acoustic simulations,'' \emph{IEEE/ACM
  Transactions on Audio, Speech, and Language Processing}, vol.~23, no.~4, pp.
  774--786, 2015.

\bibitem{fuss}
S.~Wisdom, H.~Erdogan, D.~P. Ellis, R.~Serizel, N.~Turpault, E.~Fonseca,
  J.~Salamon, P.~Seetharaman, and J.~R. Hershey, ``What’s all the fuss about
  free universal sound separation data?'' in \emph{Proc. ICASSP}.\hskip 1em
  plus 0.5em minus 0.4em\relax IEEE, 2021, pp. 186--190.

\bibitem{universal_sound_separation}
I.~Kavalerov, S.~Wisdom, H.~Erdogan, B.~Patton, K.~Wilson, J.~Le~Roux, and
  J.~R. Hershey, ``Universal sound separation,'' in \emph{Proc. WASPAA}, 2019,
  pp. 175--179.

\bibitem{conformer}
A.~Gulati, J.~Qin, C.-C. Chiu, N.~Parmar, Y.~Zhang, J.~Yu, W.~Han, S.~Wang,
  Z.~Zhang, Y.~Wu, and R.~Pang, ``{Conformer: Convolution-augmented Transformer
  for Speech Recognition},'' in \emph{Proc. {Interspeech}}, 2020, pp.
  5036--5040.

\bibitem{libricss_conformer}
S.~Chen, Y.~Wu, Z.~Chen, J.~Wu, J.~Li, T.~Yoshioka, C.~Wang, S.~Liu, and
  M.~Zhou, ``Continuous speech separation with conformer,'' in \emph{Proc.
  ICASSP}, 2021, pp. 5749--5753.

\bibitem{batchnorm}
S.~Ioffe and C.~Szegedy, ``Batch normalization: Accelerating deep network
  training by reducing internal covariate shift,'' in \emph{Proc. {ICML}},
  2015, pp. 448--456.

\bibitem{groupnorm}
Y.~Wu and K.~He, ``Group normalization,'' in \emph{Proc. ECCV}, 2018, pp.
  3--19.

\bibitem{asteroid}
M.~Pariente \emph{et~al.}, ``{Asteroid: The PyTorch-Based Audio Source
  Separation Toolkit for Researchers},'' in \emph{Proc. {Interspeech}}, 2020,
  pp. 2637--2641.

\bibitem{saijo_rccl}
K.~Saijo and T.~Ogawa, ``Remix-cycle-consistent learning on adversarially
  learned separator for accurate and stable unsupervised speech separation,''
  in \emph{Proc. ICASSP}, 2022.

\bibitem{saijo_rccl2}
------, ``Unsupervised training of sequential neural beamformer using
  coarsely-separated and non-separated signals,'' in \emph{Proc.
  {Interspeech}}, 2022, pp. 251--255.

\bibitem{adamw}
I.~Loshchilov and F.~Hutter, ``Decoupled weight decay regularization,'' in
  \emph{Proc. {ICLR}}, 2018.

\bibitem{fast_bss_eval}
R.~Scheibler, ``{SDR} --- {M}edium rare with fast computations,'' in
  \emph{Proc. ICASSP}, May 2022.

\bibitem{stoi}
C.~H. Taal, R.~C. Hendriks, R.~Heusdens, and J.~Jensen, ``An algorithm for
  intelligibility prediction of time–frequency weighted noisy speech,''
  \emph{IEEE Transactions on Audio, Speech, and Language Processing}, vol.~19,
  no.~7, pp. 2125--2136, 2011.

\bibitem{pesq}
A.~Rix, J.~Beerends, M.~Hollier, and A.~Hekstra, ``Perceptual evaluation of
  speech quality ({PESQ})-a new method for speech quality assessment of
  telephone networks and codecs,'' in \emph{Proc. ICASSP}, vol.~2, 2001, pp.
  749--752.

\bibitem{sudormrf}
E.~Tzinis, Z.~Wang, X.~Jiang, and P.~Smaragdis, ``Compute and memory efficient
  universal sound source separation,'' \emph{Journal of Signal Processing
  Systems}, vol.~94, no.~2, pp. 245--259, 2022.

\bibitem{tfgridnet}
Z.-Q. Wang, S.~Cornell, S.~Choi, Y.~Lee, B.-Y. Kim, and S.~Watanabe,
  ``Tf-gridnet: Integrating full-and sub-band modeling for speech separation,''
  \emph{arXiv preprint arXiv:2211.12433}, 2022.

\end{thebibliography}

\end{document}